\documentclass[useAMS,usenatbib]{mn2e}
\bibpunct{(}{)}{;}{a}{,}{,}
\usepackage{rotating}

\newcommand{\msun}{\mbox{M$_\odot$}}
\newcommand{\ns}{\mbox{neutron star}}

\title[Freeze-Out]{Compositional Freeze-Out of Neutron Star Crusts}
\begin{document}

\author[K. Hoffman and J. Heyl]{Kelsey Hoffman\thanks{Email: kelsey@phas.ubc.ca} and Jeremy Heyl$^{1}$\thanks{Email: heyl@phas.ubc.ca; Canada Research Chair}\\
$^{1}$Department of Physics and Astronomy, University of British Columbia \\
6224 Agricultural Road, Vancouver, British Columbia, Canada, V6T 1Z1}

\date{\today}

\pagerange{\pageref{firstpage}--\pageref{lastpage}} \pubyear{2009}

\maketitle

\label{firstpage}

\begin{abstract}
 We have investigated the crustal properties of neutron stars without fallback accretion. We have calculated the chemical evolution of the
  neutron star crust in three different cases (a modified Urca
  process without the thermal influence of a crust, a thick crust, and a direct Urca process
  with a thin crust) in order to determine the detailed composition of
  the envelope and atmosphere as the nuclear reactions freeze out.
  Using a nuclear reaction network up to technetium, we calculate the
  distribution of nuclei at various depths of the neutron star. The
  nuclear reactions quench when the cooling timescale is shorter than
  the inverse of the reaction rate.  Trace light elements among the
  calculated isotopes may have enough time to float to the surface
  before the layer crystallizes and form the atmosphere or envelope
  of the neutron star. The composition of the neutron-star envelope
  determines the total photon flux from the surface, and the
  composition of the atmosphere determines the emergent spectrum. Our
  calculations using each of the three cooling models indicate that
  without accretion of fallback the \ns\ atmospheres are dependent on the assumed 
  cooling process of the \ns. Each of the cooling methods have different
  elements composing the atmosphere: for the modified Urca process the
  atmosphere is $^{28}$Si, the thick crust has an atmosphere of
  $^{50}$Cr, and the thin crust has an atmosphere of $^{40}$Ca. In all three cases the atmospheres are composed of elements which are lighter then iron. 
\end{abstract}

\section{Introduction}

Neutron stars are the end product of the core collapse of a star with
a mass greater than 8\msun \citep{lattimer04}. A canonical neutron
star has a radius of 10 km and a mass of 1.4\msun, resulting in an
average density greater than that found in an atomic nucleus. A
neutron star is comprised of five major regions: core, inner crust,
outer crust, envelope, and atmosphere \citep{lattimer04}. The crustal
layer extends 1-2 km below the surface inward to a density of around
$10^{14}$g/cm$^3$.  Here we define the neutron-star envelope to be the
upper layer of the crust that throttles the heat flow running from a
density of $10^7{\rm g/cm}^3$ outward \citep{hernquist84,
  heylmnras01}.  The atmosphere lies at relatively low density and
comprises a column density of about 1 g/cm$^3$.

The envelope and atmosphere contain a negligible amount of the neutron-star mass
(or the mass of the crust for that matter) but play a crucial role in
shaping observations of neutron stars. In order to fully interpret the
observed emission, we need to understand the crustal composition
especially the lightest trace elements that can float to the surface
to form the envelope and atmosphere. The atmosphere shapes the
emergent spectrum, understanding the composition could yield
predictions of spectral features in the neutron star thermal
emission. The envelope, in turn, influences the transport and release
of thermal energy, a change in its composition changes the thermal
conductivity and inferred surface temperature of the neutron star
\citep{lattimer04}.

Without fallback from the supernova or other accretion, a \ns\ is
expected to have an atmosphere of iron-group
elements\citep{chiu64}. An example of a family of neutron stars which
are closest to this idealized situation are the isolated neutron
stars. The isolated \ns\ RX J185635-3754 has been studied
extensively. The surface composition of RX J185635-3754 has been
examined by fitting the x-ray spectra to various model
atmospheres, including models attributed to fallback accretion. These models have included: black body, hydrogen, helium,
iron, silicon-ash atmospheres\citep{pons02}, and later extended to two
black bodies, pure silicon, low iron silicon ash and magnetic hydrogen
atmospheres\citep{walter04}. The model atmospheres that fit the spectrum
of RX J185635-3754 the best are those with heavy elements, though the
predicted absorption lines are not seen\citep{pons02,walter04}.

Here we calculate the mass fractions that exist in the crust after a
neutron star cools long enough for the nuclear reactions to be
quenched. These mass fractions are calculated using a 489 isotope
reaction network, {\tt torch} \citep[the code is available at: {\tt
  http://cococubed.asu.edu/code\_pages/net\_torch.shtml}.][]{timmesapjs99}. The
results from the modified Urca case are compared to analytic
calculations of the freeze-out of the neutron star crust using the
nuclear statistical equilibrium software {\tt nse} (the code is
available at: {\tt http://cococubed.asu.edu/code\_pages/nse.shtml})
and the reaction rates in {\tt torch}.




\section{Neutron Star Cooling}

In order to calculate the crustal abundances three different cooling
curves were used: a modified Urca process, a thick crust, and a thin
crust. Each of the cooling curves start near a temperature of
$10^{10}$K and cools until the nuclear reactions are quenched.

In the case of the modified Urca process the nuclear reactions are
quenched before a year has past. The core temperature and the cooling
time are determined by the modified Urca equation\citep{shapbook}:
\begin{equation}
\Delta t = 1 {\rm yr} T^{-6}_9(f) \left \{ 1 -
  \left[\frac{T_9(f)}{T_9(i)}\right]^6\right\}
\label{eq:1}
\end{equation}
where $T_9(f)$ is the temperature of the outer core ($T_c$) in units
of $10^9$K. For simplicity we have assumed that the temperature of the
crust is isothermal at densities above $10^7$~g/cm$^3$
\citep[e.g.][]{heylmnras01}.

As the crust of the \ns\ cools via the diffusion of heat the to
interior, the surface of the star is expected to remain hot until the
heat reservoir of the crust has been exhausted\citep{latt94}. The
thick and the thin crust cooling models take into account this thermal
disconnect between the crust and core, whereas the modified Urca
model does not.

The cooling curve for a thick crust is based on the model in
\citet{latt94}. In this model the core is cooling rapidly and the
cooling wave takes 15 years to diffuse through the crust. This model is appropriate for a normal neutron star. The nuclear reactions are quenched before the the cooling wave passes through the crust, thus the core cooling does not affect the reactions in the crust. The cooling
curve used in this work is taken from Figure~3 in \citet{latt94},
where the hottest parts of each curve set the age and temperature of the neutron star.

The thin crust cooling curve is appropriate for a strange star. In
this case the core is cooling via the direct Urca process, where the
temperature relation is: $t\sim 20\;T^{-4}_9$s\citep{lattimer91}. The
crust of the \ns\ is cooled by crustal bremsstrahlung. The cooling of
the crust is determined from the crustal luminosity and the total
thermal energy of the crust. The luminosity of crustal bremsstrahlung
is: $L_{brems} \sim (5 \times 10^{39} {\rm erg/s})(M_{cr}/\msun)$,
where $M_{cr}$ is the mass of the \ns\ crust and the total thermal
energy of the crust is: $U_{cr} = \frac{3}{2} kT\frac{M_cr}{Am_u}$
\citep{shapbook}, where $A$ is the average atomic mass and $m_u$ is
the mass unit. The crust of a strange star reaches a density of a few
times $10^{10}\rm{g/cm^3}$, instead of neutron drip in the case of a
normal \ns\citep{stejnerPRD05}. For a crustal density of
$4.8\times10^{10}\rm{g/cm^3}$ the equilibrium nucleus is
$^{80}Zn$\citep{shapbook}. From Figure 1 of \cite{brown98} the time
scale for the cooling wave to pass through the crust is on the order
of a month. As a result, for the thin crust case the crustal cooling
is dominated by the crustal bremsstrahlung for the first month. After
a month has past the direct Urca process dominates. We take the
temperature of the crust to be
\begin{equation}
T = T_{brems}({\rm e}^{-(t/\mathrm{month})^2})+T_{DU}(1-{\rm e}^{-(t/\mathrm{month})^2})
\end{equation}
where $T_{brems}$ is the temperature of crustal bremsstrahlung and
$T_{DU}$ is the temperature at a specific time step using the equation
for direct Urca cooling. The $T_{brems}$ starts at a temperature of
$10^{10}$K and decreases by 0.1\% each step. The time, or age, of the
star is given by the crustal bremsstrahlung equation for a thin crust:
\begin{equation}
t = \frac{3}{10}k_b\frac{\msun}{Am_u}2\times10^{-31}\frac{{\rm s\; K}}{{\rm erg}} T_9^{-5}\left[1-T_{10}^5\right]
\end{equation}
for a \ns starting at a temperature of $10^{10}$K, where $T_{10}$ is
the temperature in units of $10^{10}$K.

\begin{figure}
 \includegraphics[scale=0.43]{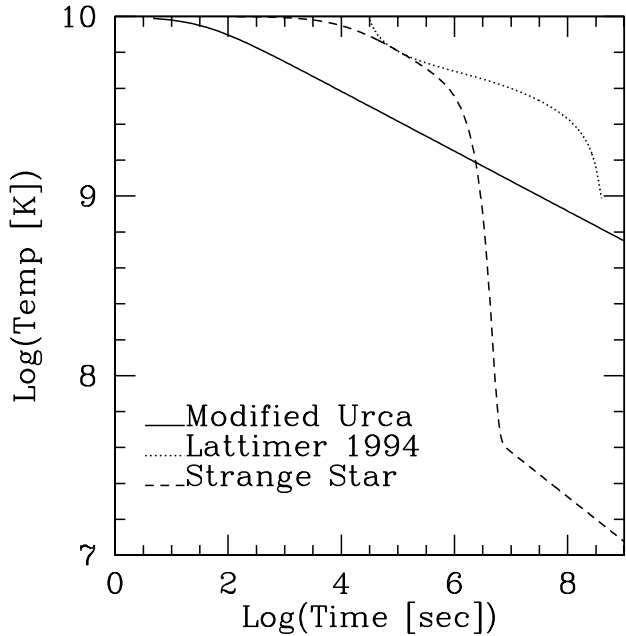}
 \caption{The three different cooling curves used for the abundance
   calculations. The methods for cooling include the modified Urca, thick crust (labelled Lattimer 1994) and a thin crust (labelled Strange Star). These curves are appropriate for
   densities at $10^7 {\rm g/cm^3}$ and above. For densities below
   $10^7{\rm g/cm^3}$, the curves would be interpolated between the
   surface temperature and the temperature at $10^7{\rm g/cm^3}$. The
   thick crust is taken from \citet{latt94}, and the thin crust is
   appropriate for a strange star.}
\label{fig:cool}
\end{figure}

The cooling curves for the 3 cases are shown in Figure~\ref{fig:cool}.
We have assumed for densities above $10^7{\rm g/cm^3}$ the crust is
isothermal and Figure \ref{fig:cool} reflects the cooling curves
appropriate for densities at $10^7{\rm g/cm^3}$ and greater. For
densities below $10^7 {\rm g/cm^3}$ the input temperature is
interpolated between the core ($T_c$) and the surface ($T_s$):
\begin{equation}
\log(T) = \left[ \frac{\log(T_c)-\log(T_s)}{7}\right]\log(\rho)+\log(T_s)
\end{equation} 
where the surface temperature is given as: $T_s = (10\; \mathrm{K}^{1/2}T_c)^{2/3}$ \citep{shapbook}.

\section{Crustal Mass Fractions}
\subsection{Running {\tt torch}}

The code {\tt torch} is a reaction network written by Frank X. Timmes
and is available on his website ({\tt http://cococubed.asu.edu}). The
software follows the abundances of 489 isotopes to technetium. The
construction of the reaction network is covered in detail in
\citet{timmesapjs00} and summarized below. The reaction network starts
by determining the mass fraction of an isotope, $i$,: $X_i =
\rho_i/\rho$ and the corresponding molar abundances of the isotope:
$Y_i = X_i/A_i$. A set of partial differential equations are
constructed from the continuity equation of the isotope:
\begin{equation}
\frac{dY_i}{dt} + \nabla \cdot(Y_i { V_i}) = \dot{R_i}
\end{equation}
where {\bf $\dot{R_i}$} is the total reaction rate and {\bf $V_i$} is
the mass diffusion velocity, which is set to zero. This results in a
set of ordinary differential equations that comprise the reaction network:
\begin{equation}
\frac{dY_i}{dt} = \dot{R}_i.
\end{equation}
The system of ordinary differential equations is integrated using a Bader-Deuflhard method coupled with MA28 sparse matrices. Three methods for integrating and eight different matrix packages are compared in \citet{timmesapjs99}.

\subsection{Mass Fraction Calculations}

For each of the densities investigated the system is started in
nuclear statistical equilibrium. The temperature and the length of
time at which the system burns is determined by the cooling curve.
After the reactions have quenched the abundances at a specific density are calculated and the isotopes which are sufficiently abundant
to compose the atmosphere or envelope are determined.

In order for the isotope to form the atmosphere it needs to have a
surface density greater than $1{\rm g/cm^2}\sim \sigma_T/m_p$, the
ratio of the Thompson cross-section to the mass of the proton. The
first step in determining which isotopes rise to the surface is to
calculate the total pressure at the density of the calculation. For
densities greater then $10^6 {\rm g/cm^3}$ the pressure is given by:
\begin{equation}
P = \frac{1.2435 \times 10^{15}}{\mu_e^{4/3}}\left(\frac{\rho}{1\rm{g/cm}^3}\right)^{4/3} \rm{dyne/cm^2},
\label{eq:2}
\end{equation}
where we have assumed that relativistic electrons dominate the
pressure and that $\mu_e=2$.  For the case where we look at a density of $10^6{\rm g/cm^3}$ the pressure is given by:
\begin{equation}
P = 1.42180\times10^{25}\phi(x) {\rm dyne/cm^2}
\end{equation}
where:
\begin{equation}
\phi(x) = [x\sqrt{1+x^2}(2x^2/3 -1) + \ln(x + \sqrt{1+x^2}\;)]
\end{equation}
and $x= p_f/m_e c$, the ratio of Fermi momentum ($p_f$) to the product
of the mass of the electron and the speed of light\citep{shapbook}.

With the pressure calculated the next step is to
calculate the column density between the particular density and the
neutron star surface: $P/g_{\rm NS}$. Where $g_{\rm NS}$ is the
surface gravity given by:
\begin{equation}
g_{\rm NS} = \frac{GM}{R^2}\left(1 - \frac{2GM}{c^2R}\right)^{-1/2}.
\end{equation}
For a 10km and 1.4\msun\ neutron star the surface gravity is
2.43$\times 10^{14}$cm/s$^2$.  Finally, the minimum mass fraction
required for an isotope to rise to form the atmosphere is given by the
ratio of 1 g/cm$^2$ to the calculated column density.

For a mass fraction sufficient to form the envelope we are looking for
isotopes which have abundances on the order of parts per million over
the entire crust, versus in the atmosphere where the isotopes need to
have abundances on the order of parts per billion or less. In order to
calculate the minimum mass fractions required for an isotope to float
to form the envelope we scale according to the column densities, we
use the typical column density of the envelope about $4 \times 10^9$
g/cm$^2$ and $1$~g/cm$^2$ for the atmosphere.

\section{Freeze-Out} \label{freezesteps}

To calculate the neutron star freeze-out we consider three different
timescales: the cooling time ($\tau_c$, for example Eq.~(\ref{eq:1}) for the modified Urca case), the
settling time ($\tau_s$), and the nuclear reaction timescale
($\tau_{\rm rxn}$). The composition of the neutron star depends on the
timescales; for the case $\tau_{\rm rxn} < \tau_c < \tau_s$, we
expect the particular species to be in nuclear statistical
equilibrium (NSE); for $\tau_c < \tau_{\rm rxn} < \tau_s$ the particular nuclear
reactions are quenched -- the star is cooling faster than the
reactions can occur; for the case $\tau_s <\tau_{\rm rxn} < \tau_c$
the isotopes can float up faster than the reactions bring 
them to NSE.  For the case $\tau_s < \tau_{\rm freeze}$
the light isotopes can float all the way up to the top before the
layer freezes.

The time when the particular layer of the neutron star freezes is
determined by comparing the potential energy between the ions that
compose the crust to their kinetic energy \citep{shapbook},
\begin{equation}
\Gamma = \frac{\mathrm{Potential \ Energy}}{\mathrm{Kinetic \ Energy}} = \frac{(Z\;e)^2}{a\;T\;k_b},
\end{equation}
where $e$ is the electron charge and $a$ is the ion radius such that the
product of $(4/3) \pi a^3$ and the ion number density is unity. When
$\Gamma > 180$ the layer freezes out, or crystallizes, and the light
species can no longer float upward.

\subsection{Settling timescale}

A second question is whether the light isotopes can reach the surface
before the layer freezes or before we look at the star.  On the other
hand the gravitational settling could be so efficient that the light
isotopes could float up before the layer cools and nuclear reactions
quench.  For these two reasons an estimate of the settling time is
crucial.

\citet{2002ApJ...574..920B} calculate the sedimentation or settling
timescale for a neutron-star atmosphere, 
\begin{equation}
\tau_s \approx 10^5 {\rm s} \left [ \frac{Z_1^{3.9}
    Z_2^{0.3} \rho_5^{1.3} }{A_1^{1.8} g_14^2 T_7^{0.3} \left
      (A_2Z_1-A_1Z_2\right )} \right ].
\end{equation}
This is an estimate of the time for a nuclide to settle down over a
pressure scale height --- a negative value means that the nuclide
ascends.  This timescale is typically around a few years for the
envelope and about $10^6$~yr for the outer crust ($\rho < 10^{12}$ g
cm$^{-3}$) for Silicon-28 in a background of Iron-56.  In particular
until the bulk of the Nickel-56 has decayed the settling time for
Silicon-28 is much larger because $A_2Z_1-A_1Z_2 \ll 1$; consequently,
the nucleons differentiate gravitationally after the nuclear reactions
effectively cease, i.e. $\tau_{\rm rxn} < \tau_s$.

\subsection{Reaction timescale}

For the modified Urca case we also calculated the reaction rate time
scales by making use of the subroutines in {\tt torch}. Each of the
rate calculations depends only on the input temperature and the
densities of the various species. As there are many different ways to
make a specific isotope, e.g $^{28}$Si, the rates which lead to the
creation of the isotope are added together to get the timescale of the
reaction rate, $\tau_{\rm rxn}$.

In order to calculate the abundance of the alpha particles and the
other species we make use of the nuclear statistical Saha equations as
implemented in {\tt nse}.  We assume that a particular species freezes
out of equilibrium when the reaction timescale exceeds the cooling
timescale.   The abundance in nuclear statistical equilibrium at the
freeze out temperature gives an alternative estimate of the final
abundance of the nuclides.


\section{Results}

In order to determine the expected composition of the \ns\ atmosphere,
in the cases of the modified Urca and the thick crust we examined 
a density of $10^7{\rm g/cm}^3$. For the thin crust a density of
$10^7{\rm g/cm^3}$ would crystallize before any of the isotopes had
time to reach the surface, so we examined a density of $10^6{\rm
  g/cm^3}$.  The results from the nuclear reaction network are
compared with those of a semi-analytic freeze-out calculation in the
modified Urca case. Each of the three cases are discussed below.

\subsection{Case 1: Modified Urca}
\subsubsection{Mass Fractions: Atmosphere}

At a density of $10^7{\rm g/cm}^3$ the corresponding pressure is:
1.1$\times 10^{24}$ {\rm dyne/cm}$^2$. At this pressure
the column density to the surface is: 4.4 $\times 10^9 {\rm
  g/cm}^2$. The resulting required minimum mass fraction required for
an isotope to be optically thick on the surface is: 2.3$\times
10^{-10}$.

\begin{figure}
 \includegraphics[scale=0.43]{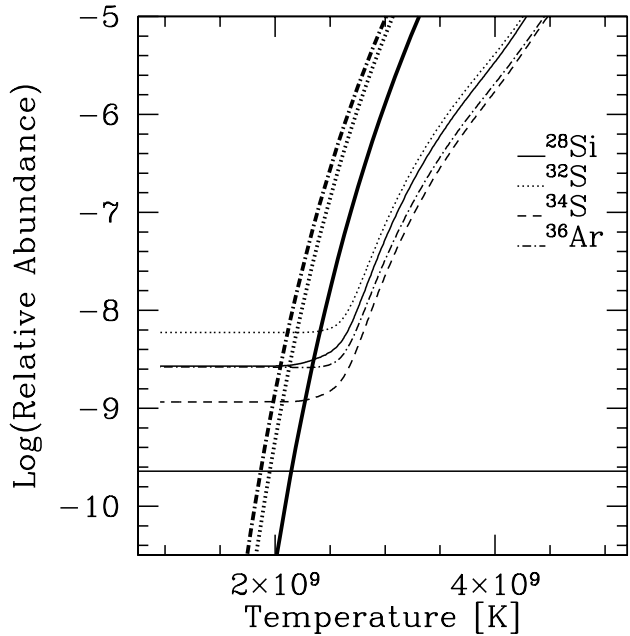}
 \caption{Lightest isotopes for which the mass fraction abundance
   would be great enough that the isotope will be optically thick on
   the neutron star surface. All of these isotopes have time to reach
   the surface before the layer crystallizes. These are the mass
   fractions for the density of $10^7 \;$g/cm$^3$. The corresponding
   pressure and column density for this neutron star density are $1.1
   \times 10^{24}\;$dyne/cm$^2$ and 4.4$\times 10^{9}$g/cm$^2$,
   respectively.  The horizontal line indicates the minimum abundance
   required for an isotope to have a surface density of 1g/cm$^2$.
   The abundances for $^{28}$Si, $^{32}$S, and $^{36}$Ar in nuclear
   statistical equilibrium using {\tt nse} are depicted for comparison
   by the nearly vertical, heavy lines.  }
\label{fig:surf7}
\label{fig:tde}
\end{figure}

Isotopes with a mass fraction greater than 2.3$\times 10^{-10}$ will
have a surface density of 1${\rm g/cm}^2$. The lightest elements to be
optically thick on the surface and have time to reach the surface
before crystallization of the layer occurs are shown in Figure
\ref{fig:surf7}, where the horizontal line indicates the minimum mass
fraction required to be optically thick on the surface. The lightest elements to
rise to the surface which are optically thick are: $^{28}$Si, $^{32}$S,
$^{34}$S, and $^{36}$Ar.  In particular the abundance of $^{28}$Si is
about $3\times 10^{-9}$ so a layer $\sim 10$g/cm$^{3}$ of silicon lies
on the surface of the star.

\subsubsection{Freeze-Out}

\begin{figure}
\includegraphics[scale=0.43]{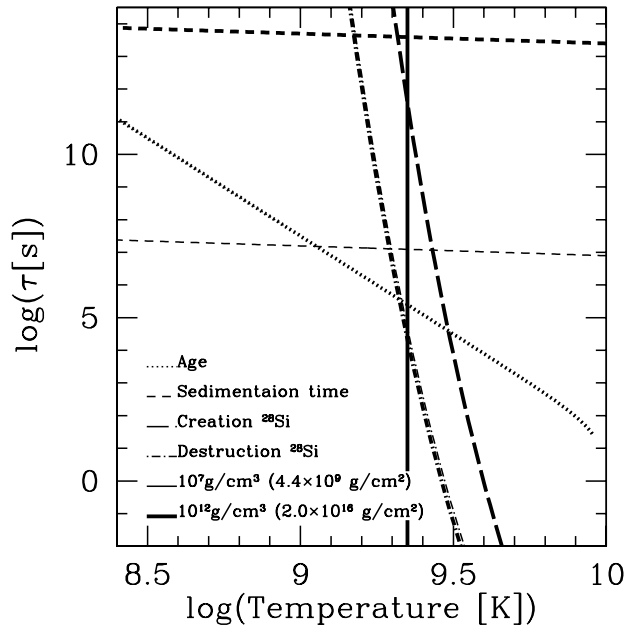}
\caption{Comparing the cooling, settling, and nuclear reaction
  timescales of $^{28}$Si for the neutron star densities of $10^7$
  g/cm$^3$ and $10^{12}$g/cm$^3$. The timescales for the density of
  $10^{12}{\rm g/cm}^3$ are denoted by the heavier line, while the
  thinner line is for the density of $10^7{\rm g/cm}^3$. The heavy
  vertical line denotes the temperature at which the layer at
  $10^{12}{\rm g/cm^3}$ crystallizes, the temperature for the
  crystallization of the layer at $10^7{\rm g/cm^3}$ is cooler than
  $10^{8.5}$~K and so lies to the left of the plot. The layer of
  $10^{12}{\rm g/cm^3}$ crystallizes at $2.2\times 10^9$K, and the
  layer at $10^7 {\rm g/cm^3}$ crystallizes at $4.8 \times 10^7$K. The
  numbers in the parentheses give the corresponding surface density at
  the two neutron-star number densities. Note that the cooling and
  reaction timescales overlap for the two densities.}
\label{fig:timfreeze}
\end{figure}

Using the steps outlined in \S \ref{freezesteps} we have
calculated the cooling ($\tau_c$), settling ($\tau_s$), and the
nuclear reaction ($\tau_{\rm rxn}$) timescales for the case of
$^{28}$Si. The results of these calculations are displayed in Figure
\ref{fig:timfreeze}. These calculations compare the age of the neutron
star to the settling, crystallization temperature, creation and
destruction timescales of $^{28}$Si for two densities: $10^7 \;
\rm{g/cm}^3$ and $10^{12} \; \rm{g/cm}^3$. The temperatures at which
the layers crystallize are $4.8\times 10^7$K and $2.2\times 10^9$K,
for the densities of $10^7{\rm g/cm^3}$ and $10^{12}{\rm g/cm^3}$,
respectively.  The creation and destruction rates are $d\ln X_i/dt$,
where $X_i$ is the abundance of the $^{28}$Si isotope. These rates are
calculated by using the routines in {\tt torch} to determine the
energy release per unit mass, these are then multiplied by the
abundances calculated from the {\tt nse} code. The abundances were
also calculated using the nuclear Saha equation and are the output of
the {\tt nse} routine. The two different methods for calculating the
relative abundances, the output from {\tt torch} and using {\tt nse}
are displayed in Figure \ref{fig:tde}. It is clear that the results
from {\tt torch} do not follow nuclear statistical equilibrium
precisely.


Using Figure~\ref{fig:timfreeze} we can find the quenching temperature
for the silicon reactions. The reactions become quenched when the
reaction rate time becomes longer than the age of the neutron star.
From the plot the quenching temperature for the reactions which
destroy silicon are $2.12\times10^9 \rm{K}$ and $2.15 \times 10^9
\rm{K}$ for the densities of $10^{12}\rm{g/cm}^3$ and
$10^{7}\rm{g/cm}^3$, respectively. The reactions which create silicon
are quenched at $3.05\times 10^9 \rm{K}$ for both of the
densities.   These approximately give the temperature range over which
the abundance of silicon levels out in Figure~\ref{fig:tde}.
The relative abundances of $^{28}$Si at the quenching
temperatures for the destructive reactions are $4.86\times10^{-13}$
and $2.36\times10^{-10}$, for densities of $10^{12}\rm{g/cm}^3$ and
$10^{7}\rm{g/cm}^3$, respectively. The relative abundances of silicon
for the creation reactions are: $5.82\times10^{-9}$ at a density of
$10^{12}\rm{g/cm}^3$ and $1.84\times 10^{-6}$ at a density of
$10^7\rm{g/cm}^3$.



\subsection{Case 2: Thick Crust}

\begin{figure}
\includegraphics[scale=0.43]{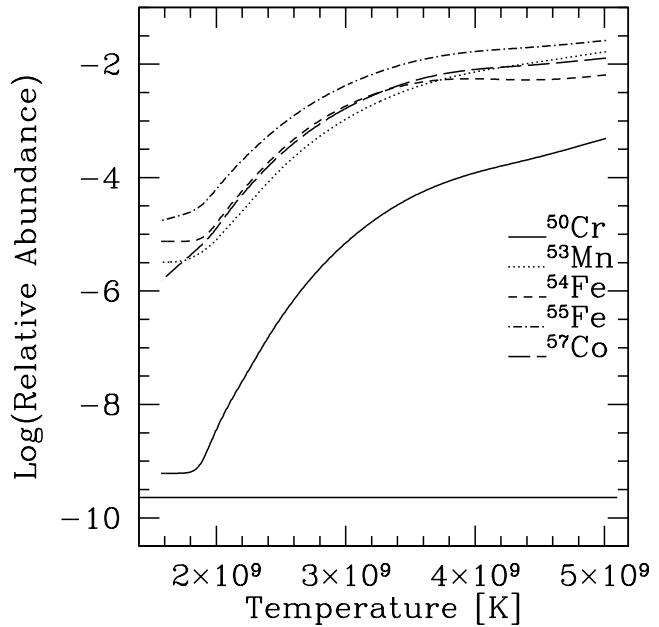}
\caption{The lightest isotopes for which the mass fraction abundances
  are large enough that the isotopes which can reach the surface are
  optically thick. These are the mass fractions for a density of
  $10^7\rm{g/cm^3}$ and for a neutron star with a thick crust. The
  horizontal line indicates the minimum abundance required for an
  isotope to have a surface density of $1{\rm g/cm^2}$.}
\label{fig:latt94}
\end{figure}

In the case of the thick crust, the density of $10^7{\rm g/cm^3}$ was
examined for the isotopes which can rise to the surface. The pressure,
column density to the surface and the minimum mass fraction required
to be optically thick are the same for the modified Urca case. For the
thick crust the lightest isotopes which are optically thick and can
rise to the surface are $^{50}$Cr, $^{53}$Mn, $^{54}$Fe, $^{55}$Fe,
and $^{57}$Co, as shown in Figure \ref{fig:latt94}.

\subsection{Case 3: Thin Crust}

\begin{figure}
\includegraphics[scale=0.43]{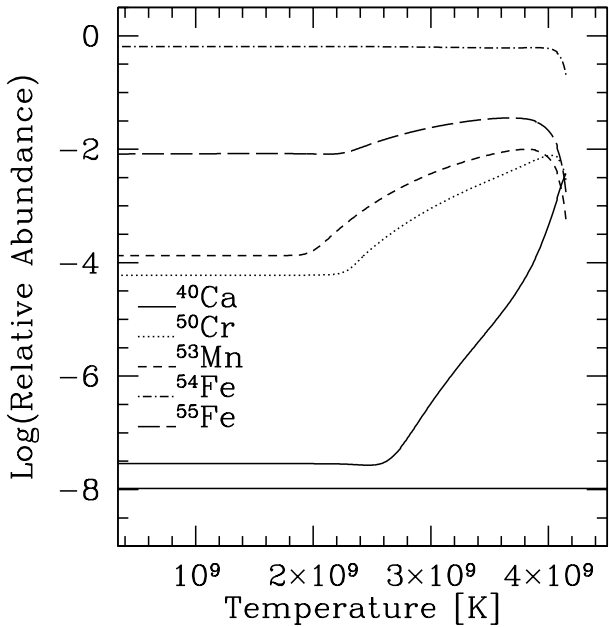}
\caption{The lightest isotopes which can reach the surface and are
  also optically thick. These isotopes are for a density of
  $10^6\rm{g/cm^3}$ for a neutron star with a thin crust. The
  corresponding pressure and column density at this density are
  $2.3\times10^{22}{\rm dyne/cm^2}$ and $9.6\times10^7{\rm g/cm^2}$,
  respectively. The horizontal line indicates the minimum mass
  fraction required in order for an isotope to be optically think. }
\label{fig:sqm6}
\end{figure}

For the thin crust the isotopes which had the possibility to be
optically thick could not reach the surface before the layer
crystallized for densities at $10^7\mathrm{g/cm}^3$ and higher;
consequently, we examined the layer at a density of
$10^6\mathrm{g/cm}^3$. At this density the corresponding pressure is:
$2.3\times10^{22}\mathrm{dyne/cm}^2$. The column density to the
surface at this pressure is $9.6\times10^7\mathrm{g/cm}^2$, requiring
a minimum mass fraction of $1.0\times 10^{-8}$ for an isotope to
optically thick on the surface. The isotopes which are not only
optically thick, but can also rise to the surface are shown in Figure
\ref{fig:sqm6}. These isotopes include $^{40}$Ca, $^{50}$Cr,
$^{53}$Mn, $^{54}$Fe, and $^{55}$Fe.

\section{Conclusions}

In the case of cooling via the modified Urca process, without the thermal influence of a crust,  our results
(e.g. Fig.~\ref{fig:timfreeze}) show that silicon has sufficient time
to float to the top from a density of $10^7$~g/cm$^3$ before the layer freezes or
we observe it.  However, the settling time from $10^{12}$~g/cm$^3$ is
too long for light species to float up before the crust freezes;
therefore, we can conclude that the atmosphere in this case is likely
to be composed of silicon but the envelope is likely to be composed of
iron-group elements that have been chemically separated by
gravitational settling.  Deeper layers are unlikely be chemically
separated at least by gravity.

We have used the {\tt torch} code in order to calculate the lightest
isotopes which would rise to the neutron star surface and have
compared these results to semi-analytic calculations. Using the {\tt
  torch} code we calculated the the mass fractions at a density of
$10^7 {\rm g/cm}^3$.  We found the lightest isotopes to rise to the
surface and be optically thick in the atmosphere are: $^{28}$Si,
$^{30}$Si, $^{31}$P, $^{32}$S, $^{33}$S, $^{34}$S, and $^{36}$Ar. On
the other hand, there is not sufficient time for light elements to
percolate from high density to form the envelope.  However,
calculations of freeze out at higher densities are required to
determine the precise composition of the envelope.

We also did semi-analytic calculations of the freeze-out of $^{28}$Si,
for the modified Urca case, assuming local thermodynamic equilibrium
until the cooling rate exceeds the reaction rate. We used the rates
from {\tt torch} and the abundances from the code {\tt nse} in order
to calculate the rates of nuclear reactions involving $^{28}$Si; the
reactions become quenched when the reaction time is longer than the
age of the neutron star. We found the creation reactions are quenched
at a temperature of $3.05\times 10^9$K for both the densities of
$10^7\mathrm{g/cm}^3$ and $10^{12} \mathrm{g/cm}^3$. The reactions which
destroy silicon are quenched at $2.12\times 10^9$K and $2.15\times
10^9$K at a density of $10^{12} \mathrm{g/cm}^3$ and $10^7 \mathrm{g/cm}^3$,
respectively. The calculated quenching temperatures of the silicon
reactions agree with the results calculated using the {\tt torch} code
directly.

In the case of the thick crust, the atmosphere could be formed by
$^{50}$Cr rising to the surface from a density of $10^7\mathrm{
  g/cm^3}$. For a \ns\ with a thin crust and direct Urca cooling the
layers at $10^7\mathrm{ g/cm^3}$ and higher crystallize before the
isotopes could have time to reach the surface. The atmosphere in the
thin crust case could be formed from $^{40}$Ca rising to the surface
from a layer at a density of $10^6\mathrm{g/cm^3}$.

Unless there has been significant accretion either from the supernova
debris, the interstellar medium or a companion, neutron-star
atmospheres are unlikely to be composed of iron, helium or hydrogen.
The type of isotope composing the atmosphere depends on the cooling mechanism at work in 
the star: $^{28}$Si for modified Urca, $^{50}$ Cr for a thick crust,
and $^{40}$Ca for a thin crust and direct Urca cooling. The formation of these atmospheres can provide additional justification, with fallback accretion,  for isolated neutron stars fit with intermediate atmosphere models. Understanding
how this novel composition of the atmospheres affects the neutron-star
emission may provide new insights on the observed spectra of neutron
stars.

\section*{Acknowledgements}

K.H. is supported by a graduate fellowship from the Natural Sciences
and Engineering Research Council of Canada.  The Natural Sciences and
Engineering Research Council of Canada, Canadian Foundation for
Innovation and the British Columbia Knowledge Development Fund
supported this work.  Correspondence and requests for materials should
be addressed to heyl@phas.ubc.ca.  This research has made use of
NASA's Astrophysics Data System Bibliographic Services.

\bibliographystyle{mn2e}
\bibliography{complete}
\label{lastpage}
\end{document}